\documentclass[a4paper]{article}
\usepackage{amssymb,amsmath}
\usepackage[dvips]{graphics,epsfig}
\usepackage{floatflt}

\def\ka{\hbox{\ae}}
\begin{document}
\hoffset=-25mm
\voffset=-20mm
\title{\bf Sub-Doppler resolution with double coherently driving fields}
\author{Po Dong$^{{1}}$\footnote{E-mail: scip9544@nus.edu.sg},
A.K. Popov$^{\text{2}}$\footnote{E-mail: popov@ksc.krasn.ru},
Tang Sing Hai$^1$, and Jin-Yue Gao$^3$\\ 
$^1$Physics Department, National University of Singapore, 119260, Singapore\\
$^2$Institute for Physics of Russian Academy of Sciences and Krasnoyarsk State University,\\
  660036 Krasnoyarsk, Russia\\
$^3$Physics Department, Jilin University, Changchun, Jilin 130023, China
}
\date{}
\maketitle

\begin{abstract}
We propose a four-level model where sub-Doppler resolution as
well as enhanced absorption of a weak probe field are realized by
using two coherently driving fields. We show that spectral
resolution can be improved by modifying the coherent fields
intensity and frequencies.
\end{abstract}
PACS number(s): 42.50.Hz,
42.50.Ct,
42.50.Gy
\section{Introduction}
Multi-photon transitions are embedded in a variety of optical
effects based on correlated absorption and emission of two or
more photons. Step-wise  and multi-photon processes can be
distinguished by their frequency-correlation properties
\cite{Rau,Vved}. Sub-Doppler resolution for an inhomogeneously
broadened medium based on multi-photon processes has attracted
much attention in recent decades. The physical essence for this
method lies in the fact that a detuning for multi-photon
transition stipulated by Doppler frequency shift can be reduced
or eliminated by adopting appropriate light propagation
direction, since the detuning is the sum or difference of
multiple single-photon transition detunings. Sub-Doppler
nonlinear optical resonance as appearance of quantum coherence
and interference in the context of frequency-correlation
properties  of the coherent components have been proposed in
\cite{Feok}. A widely used method is Doppler-free two- or
multi-photon absorption  where the velocity dependent detuning
could be removed in the case that the wave vectors of the
interacting beams sum down to zero (for a review, see
\cite{Let}). This type of sub-Doppler spectrum is characterized
by a large detuning from the intermediate resonance, measuring
the fluorescence from the upper level as well as by small cross
sections for these multi-photon processes. Enhanced by
intermediate resonance sub-Doppler processes based on the use of
strong driving fields and on a change of frequency-correlation
properties of resonant multi-photon process in strong resonant
fields, have been proposed in \cite{Feok} and further developed
in \cite{Coh,Tal} (also in \cite{Vved}). Substantial enhancement
in absorption, gain and fluorescence controlled with the auxiliary
appropriately propagating electro-magnetic wave has been
predicted. An interference nature of the spectrum modification
has been stressed, which implies that along the growth of the
absorption (gain) in certain spectral intervals, integral over
the frequency may even decrease. A role of increase of intensity
of a probe field as well as features of Doppler-free lasers were
explored too.

Control of atomic response with intense coupling lasers has been
a subject of many intensive studies in the context of
electromagnetically induced transparency (EIT) \cite{S.E.
Harris}, amplification without inversion (AWI)
\cite{AWI1,AWI2,Kuch,Spi}, enhancement of the refractive index
without absorption \cite{M. O. Scully} and so on (for review see
\cite{AWI1,Spi,Rev}). Sub-Doppler resolution by using intense
coherently driving field(s) at other transition(s) has been
recently further explored in \cite{Vemuri,YiFu Zhu}. In Ref.
\cite{Vemuri} probe weak field absorption spectrum in three-level
schemes was considered. It was pointed out that the linewidth of
one of the two Autler-Townes absorption peaks can be reduced by
match of the coupling field intensity and frequency. In a later
work \cite{YiFu Zhu}, an experimental observation of
sub-Doppler linewidth in a Doppler-broadened $\Lambda $-type Rb
atomic system was reported.  Related works to reduce Doppler broadening
with atomic coherence effects could be found in
Ref. \cite{Wang DZ} and multiple sub-Doppler lines have been
shown to achieve with a strong coupling field and a saturation
effects in a three-level system\cite{Dong1}.

In the recent papers \cite{Popov}, cancellation of Doppler
broadening by applying different frequency fields was proposed
with one or two coherently driving lasers. As it has been
outlined, sub-Doppler structures can be induced without
population redistribution. At that not only absorption and gain,
but four-wave mixing output can be enhanced through Doppler-free
coupling controlled with auxiliary co- or counter-propagating
driving electromagnetic radiation.

In this paper, we propose a four-level scheme, where both upper
and lower levels of a probe transition are coupled to other
levels by strong coherent fields. In this scheme, four
absorption peaks could be found on account of the fact that both
the upper and lower states are split into Autler-Townes doublets.
Sub-Doppler resolution is achievable because the modified
two-photon transition occurs contributing in the probe process and
two-photon detunings, resulting from the Doppler shift, could be
reduced by choosing appropriate optical geometry. A crucial role
of the ratio of the frequencies of the coupled transitions is
outlined. The results obtained on the basis of the developed
theory are accompanied with numerical simulation addressed to
real experimental schemes. Despite the decreased integral
intensity, we predict enhanced laser-induced sub-Doppler
absorption peaks along with transparency windows. The features
are interpreted in the terms of quantum coherence and
interference processes.

The paper is organized into four sections. In the next section,
we present a four-level model and the density-matrix equations
describing the system. The solution in the limit of a weak probe
and steady-state condition is obtained and the absorption
spectrum is then analyzed, first without Doppler broadening. In
Sec. III, taking into consideration Doppler broadening, we
demonstrate that the multiple sub-Doppler lines as well as
enhanced absorption at the resonance could be obtained in this
scheme. From numerical illustrations of the effect in a practical
medium, the conditions for and features of Doppler-free
resonance, stipulated by compensation of Doppler shifts with
light shifts are discussed. In Sec. IV, we summarize the results.

\section{Model and absorption spectrum at homogeneously broadened
probe transition}
\begin{floatingfigure}{50mm}
\includegraphics[width=45mm]{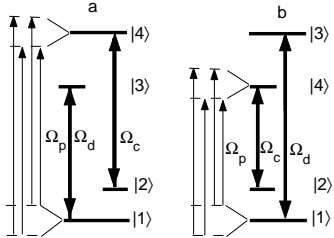}
\caption{\label{f1}\footnotesize The energy level schemes under consideration.}
\end{floatingfigure}

We consider a closed four-level scheme shown in Fig.\ref{f1}(a). In this scheme two
coherent driving fields $E_c$ and $E_d$ with coupling Rabi frequencies $\Omega _c=-{\bf
E_c\cdot d_{13}}/2\hbar$ and $\Omega _d=-{\bf E_d\cdot d_{24}}/2\hbar$ interact with
the transitions labeled $\left| 4\right\rangle -\left| 2\right\rangle $ and $\left|
3\right\rangle -\left| 1\right\rangle $, respectively. The transition $\left|
4\right\rangle -\left| 1\right\rangle $ is probed by the weak nonperturbating field
$E_p$. Absorption index $\alpha_p$ for this field, reduced by it's  value
$\alpha_{p0}^0$ at $\omega_p= \omega_{41}$, $\Omega _c=\Omega _d=0$ is found as
$\alpha_p/\alpha_{p0}^0=Re\{\chi_p/\chi_{p0}^0\}$, where $\chi_p$ is corresponding
dressed susceptibility. The later is convenient to calculate with aid of dressed
nonlinear polarization $ P^{NL}$ and density matrix $\rho _{ij}$ as $ P^{NL}(\omega
_{p})=N \chi_{p}E_{p}/2=N\rho _{ij}{d}_{ji}$ ($d_{ji}$ is transition electrodipole
moment, N - number density of atoms). In the framework of semi- classical theory and
using the standard density matrix formalism with the rotating wave approximation, the
description equations of this scheme can be written in general form as:
\begin{eqnarray}\label{ro}
&L_{nn}\rho_{nn}=q_n-i[V, \rho]_{nn}+\sum_{m>n}
\gamma_{mn}\rho_{mm}; L_{14}\rho_{14}=L_p\rho_p=-i[V,\rho]_{14}\
(etc.),&
\end{eqnarray}
where $L_{ij}=d/dt\,+\Gamma_{ij};\quad
V_{14}=\Omega_{p}\cdot\exp\{i\Delta_pt\};\quad \Omega_{p}=-{\bf
E_p\cdot d_{14}}/2\hbar$; $ \Delta _ p = \omega _ p-\omega_ {41}
$\ (etc.) are frequency detunings from the corresponding
resonance; $ \Gamma _ {ij} $ - homogeneous half-widths of
transitions (in absence of collisions $ \Gamma _ {mn} = (\Gamma_m
+ \Gamma_n)/2$); $ \Gamma _ n = \sum _ j\gamma _ {nj} $ - inverse
lifetimes of levels; $ \gamma _ {mn} $ - rate of relaxation from
the level $m$ to $n$, $ q _ n = \sum _ j w _ {nj} r _ j $ - rate
of incoherent exitation to a state $ n $ from the underlying
levels.

We represent density matrix elements as $\rho_{14} =
r_p\cdot\exp\{i\Delta_pt\}, \rho_{33}=r_3$ etc. Then in a
steady-state regime a set of density-matrix equation  may be
reduced to the set of algebraic equations for the amplitudes
$r_{ij}$. Analytical solution of this system of 10 coupled
equation both for closed and open four-level scheme is given in
\cite{Spi,Mys}. In the case under consideration incoherent
exitation of the upper levels as well as relaxation from the
level 3 to 2 are supposed to be negligible small. Then $r_{2}=
r_{4}= r_c=0$ and equation system (\ref{ro}) reduces to
\begin{eqnarray}\label{ro1}
&P_pr_p=i\Omega_pr_1+ir_{12}\Omega_c-i\Omega_dr_{34},\
P_{12}r_{12}=ir_p\Omega_c^{*}-i\Omega_dr_{32},&\label{1}\\
&P_{34}r_{34}=-i\Omega_d^{*}r_p+ir_{d}^{*}\Omega_p+ir_{32}\Omega_c,\
P_{23}r_{23}=-i\Omega_c r_{43}+ir_{21}\Omega_d.&\label{2}\\
&P_{d}r_{d}=i\Omega_d(r_1-r_3),\ \Gamma_3r_3=
-2Re\{i\Omega_d^*r_d\},\  r_{1}= 1 - r_{3}.&
\end{eqnarray}
Here $ P _ {14,13}\equiv P _ {p,d} = \Gamma _{p,d} + i\Delta
_{p,d}$, $ P _ {12} = \Gamma _{12} + i(\Delta _ p -\Delta _ c)$,
$ P _ {34} =  \Gamma _{34} + i(\Delta _ p -\Delta _ d)$, $ P _
{23} =  \Gamma _{23} + i(\Delta_c-\Delta _ p+\Delta _ d)$.  For
atom moving with speed $\rm v $, Doppler shift of resonances must be
taken into account by substituting  $ \Delta _ j $ for $ \Delta _
j ^ {'} = \Delta _ j - {\bf k} _ j{\bf v} $.

Under considered approximation solution for two level system
$\left| 1\right\rangle - \left|3 \right\rangle$ is found apart of
the other elements.
\begin{eqnarray}\label{ro2}
&r_d=i\dfrac{\Omega_dP_d^*}{\Gamma_d^2(1+\ka_d)+\Delta_d^2},\
r_3=\dfrac{\Gamma_d^2\ka/2}{\Gamma_d^2(1+\ka)+\Delta_d^2},\
\ka=\dfrac{4|\Omega_d|^2}{\Gamma_3\Gamma_d}.&
\end{eqnarray}
Dressed susceptibility for the probe field is found in the
notations of \cite{Spi,Mys} as:
\begin{eqnarray}\label{alpha}
&\dfrac{\chi_p}{\chi_{p0}^0}= \dfrac{\Gamma_p}{P_p}R_p,\
R_p=\dfrac{r_1(1+g_5+v_5)-(r_1-r_3)(1+g_5-v_6)g_1}{1+g_5+v_5+g_4(1+g_5-v_6)+v_4(1+v_5-g_6)},
&
\end{eqnarray}
$$g_1=\dfrac{|\Omega_d|^2}{P_dP_{34}},
g_4=\dfrac{|\Omega_d|^2}{P_pP_{34}},
g_5=\dfrac{|\Omega_d|^2}{P_{12}P_{23}^*},
g_6=\dfrac{|\Omega_d|^2}{P_{34}P_{23}^*},
v_4=\dfrac{|\Omega_c|^2}{P_pP_{12}},
v_5=\dfrac{|\Omega_c|^2}{P_{34}P_{23}^*},
v_6=\dfrac{|\Omega_c|^2}{P_{12}P_{23}^*}.$$
At $\Omega _d=0$ all $g_i=0$ and the equation
(\ref{alpha}) reduces to that describing $\Lambda$ scheme:
\begin{eqnarray}
&\dfrac{\alpha_p}{\alpha_{0p}^0}=Re\dfrac{\Gamma_p[\Gamma_{12}+i(\Delta_p-\Delta_c)]}
{(\Gamma_p+i\Delta_p)[\Gamma_{12}+i(\Delta_p-\Delta_c)]+|\Omega_{c}|^2}
=-Re \dfrac{\Gamma_p[\Gamma_{12}+i(\Delta_p-\Delta_c)]}
{(\Delta_p-\delta_1)(\Delta_p-\delta_2)},&\label{spl}\\
& \delta_{1,2}=
\dfrac{\Delta_c+i(\Gamma_{12}+\Gamma_p)}{2}\mp\sqrt{\dfrac{\Delta_c+i(\Gamma_{12}-\Gamma_p)}{2}
+|\Omega_c|^2}.&\label{rt}
\end{eqnarray}

Following to \cite{Feok}, we introduce frequency-correlation
factor
\begin{eqnarray}
&M_{1,2}=\dfrac{d\delta_{1,2}}{d\Delta_c}=\dfrac{1}{2}\left[1\mp\dfrac{\Delta_c}
{\sqrt{4|\Omega_c|^2+\Delta_c^2}}\right].&\label{m}
\end{eqnarray}
The denominator in (\ref{spl}) displays two resonances. At
$\Omega_c\rightarrow 0$ we obtain $\delta_{1}\rightarrow
0+i\Gamma_p, \delta_{2} \rightarrow \Delta_c+i\Gamma_{12}$. This
indicates one resonance at $\Delta_p=0$, the HWHM is $\Gamma_p$,
which corresponds to one-photon resonance with no correlation with
$\omega_c$ ($M_1=d\delta_1/d\Delta_c=0$). The second resonance at
$\Delta_p=\Delta_c$ is of HWHM $\Gamma_{12}$ that corresponds to
two-photon resonance, fully correlated with $\omega_c$
($M_2=d\delta_2/d\Delta_c=1$). With growth of the coupling Rabi
frequency $\Omega_c$ the resonance becomes split in two
component, their frequency-correlation properties modify, and
$M_1\approx M_2\approx 1/2$ at $|\Omega_c|^2\gg\{|\Delta_c|^2,
\Gamma_{p,12}^2\}$, which neither correspond to one- nor to
two-photon processes. HWHM of the resonances becomes also nearly
equal to each other and to $(\Gamma_{12}+\Gamma_p)/2$. Note that
all the effects are determined by the coherence $\rho_{12}$ induced in
the transition $\left| 1\right\rangle - \left|2 \right\rangle$ by
two coupled fields,
and that always $M_1+M_2=1$. More detailed discussion can be
found in \cite{Rau,Vved,Feok,Popov}.

In the alternative case $\Omega_c=0$ all $v_i=0$ and the equation
(\ref{alpha}) reduces to that describing $V$ scheme:
\begin{eqnarray}
&\dfrac{\alpha_p}{\alpha_{p0}^0}=Re\left\{\Gamma_p
\dfrac{[\Gamma_{34}+i(\Delta_p-\Delta_d)]-
(r_1-r_3)|\Omega_d|^2/(\Gamma_p+i\Delta_p)}
{(\Gamma_p+i\Delta_p)[\Gamma_{34}+i(\Delta_p-\Delta_d)]+|\Omega_{d}|^2}
\right\}.&\label{awi}
\end{eqnarray}
The structure of the denominator is similar to (\ref{spl}) and
determined by the coherence $\rho_{34}$. Additional term in the
nominator is stipulated by the coherence induced at the
transition $\left| 1\right\rangle - \left|3 \right\rangle$ with
not zero populations of the levels unlike the transition
$\left|2\right\rangle - \left|4 \right\rangle$. Alongside with the
coherence $\rho_{34}$ this term is a
source of the nonlinear interference effects (NIEF) in absorption
(gain) and refraction. Specific features of NIEF in coupled
Doppler broadened $\Lambda$, $V$ and ladder schemes were explored
in \cite{Vved,NIEF}. It has been outlined that frequency
integrated absorption index is proportional to  $r_1-r_4$ only,
i.e. it's change is determined only by the population change. As
it was first outlined in \cite{AWI1,NIEF}, indeed NIEF give rise
to difference in the line shapes of pure absorption and emission
spectra. As the consequence of this effect, the appearance of
amplification without inversion on the base of NIEF has been
predicted.  Amplification without inversion was introduced and its features
were explicitly analyzed and illustrated for the model of neon transitions
in the early publication \cite{AWI1}.

Specific feature of the case under consideration in this paper
is that both of the levels of the probe transition can be driven
independently. This gives rise to multiple resonance structure,
corresponding to the roots of denominator in (\ref{alpha}). If
resonance splitting is much greater than their widths, we can set
$\Gamma_{ij}=0$,  and the resonance values of $\Delta_p$ are
described by the equation:
\begin{eqnarray}
&[\Delta_p-\Delta_d-\Delta_c][\Delta_p(\Delta_p-\Delta_d)(\Delta_p-\Delta_c)-
(\Delta_p-\Delta_c)\Omega_d^2-(\Delta_p-\Delta_d)\Omega_c^2]-&\nonumber\\
&-\Delta_p[(\Delta_p-\Delta_d)\Omega_d^2+(\Delta_p-\Delta_c)\Omega_c^2]
+(\Omega_d^2-\Omega_c^2)^2=0.&\label{rt1}
\end{eqnarray}
Basically, this equation possesses four roots, that indicates
appearance of four nonlinear resonances, determined by splitting
of each of the levels $\left|1\right\rangle$ and $\left|4
\right\rangle$ into two quasi levels.

In the case $\Delta_d=-\Delta_c=\Delta$ resonance positions are
given by the equation:
\begin{eqnarray}
&\Delta_p^2=(\Delta^2+2\Omega_d^2+2\Omega_c^2)/2\pm
\sqrt{[(\Delta^2+2\Omega_d^2+2\Omega_c^2)/2]^2-(\Omega_d^2-\Omega_c^2)^2}.&\label{rt2}
\end{eqnarray}
At $|\Omega_d|=|\Omega_c|=|\Omega|$ two resonances merge in one
not shifted resonance at $\Delta_p^{(1)}=0$, the other two
resonances are given by
\begin{eqnarray}
&\Delta_p^{(2,3)}=\pm\sqrt{\Delta^2+4\Omega^2}.&\label{rt3}
\end{eqnarray}

In the case $\Delta_d=\Delta_c=0$ resonance positions are found
as:
\begin{eqnarray}
&\Delta_p^{(1,2)}=\pm(|\Omega_d|-|\Omega_c|);\
\Delta_p^{(3,4)}=\pm(|\Omega_d|+|\Omega_c|).&\label{rt4}
\end{eqnarray}
\begin{figure}[!h]
  \centering
\includegraphics[width=.3\textwidth]{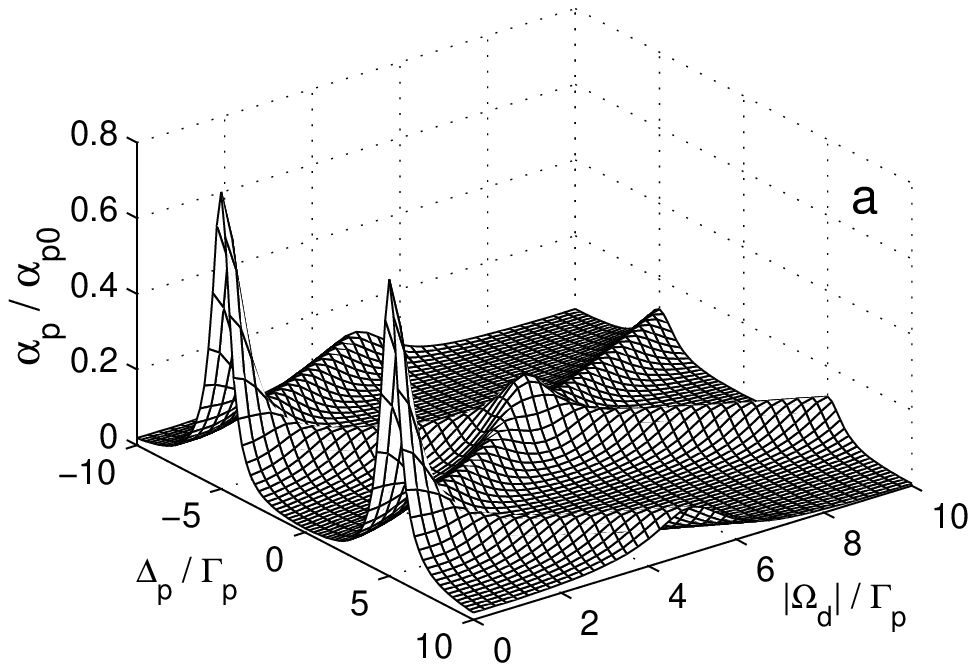}
\includegraphics[width=.3\textwidth]{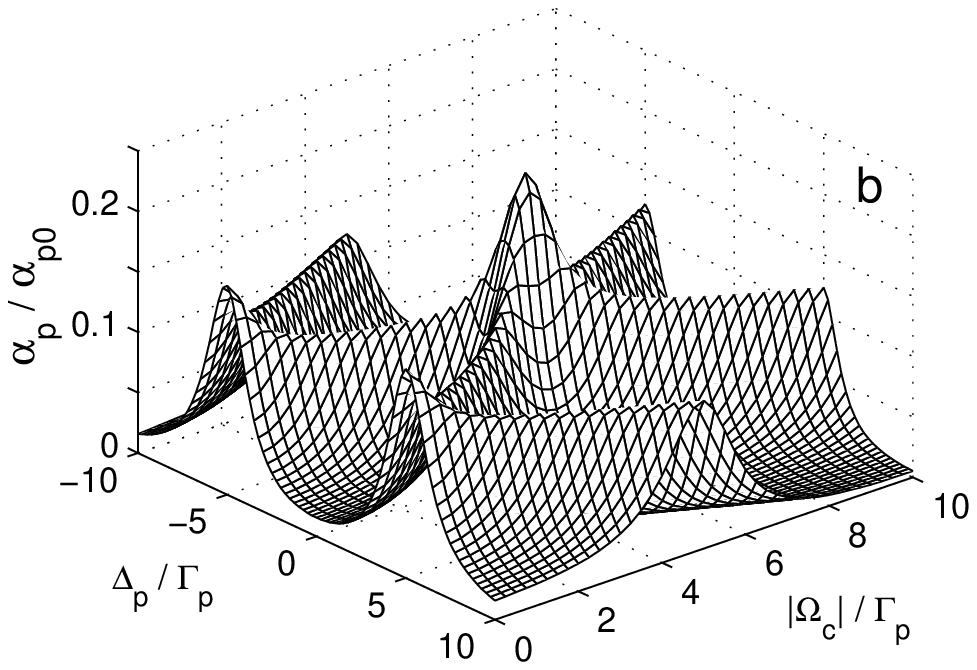}
\includegraphics[width=.3\textwidth]{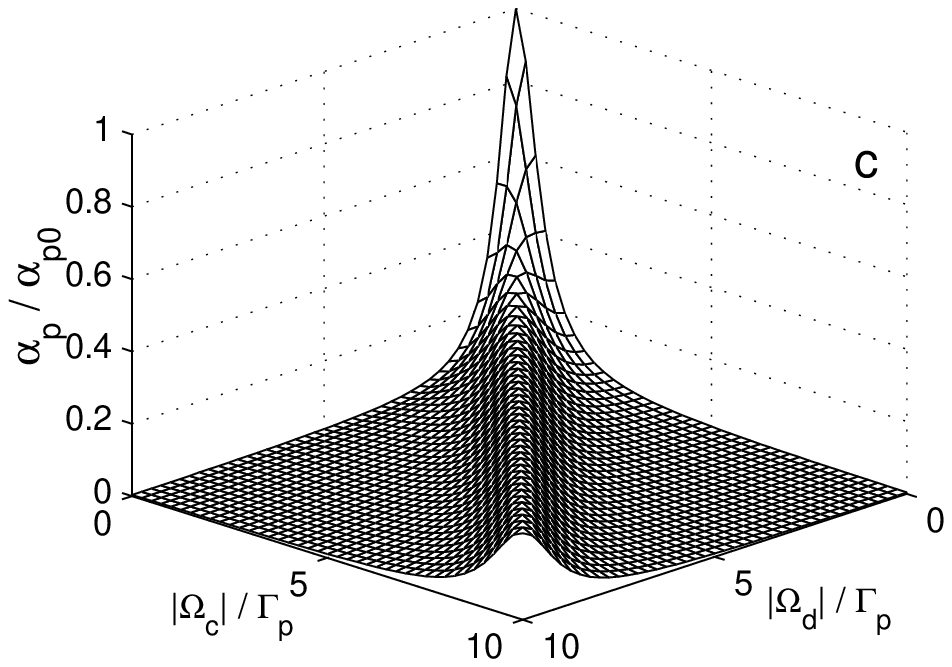}
  \caption{\footnotesize Absorption spectrum at homogeneously broadened probe
transition $\left| 1\right\rangle -\left| 4\right\rangle $
controlled with two driving field. The detunings and  Rabi frequencies
of the driving fields are:
$\Delta_c=\Delta_d=0$. (a) $\left|\Omega _c\right|=5\Gamma _{p}$.
(b) $\left|\Omega _d\right|=5\Gamma _{p}$. (c) $\Delta_p=0$.}\label{f2}
\end{figure}
\begin{floatingfigure}{75mm}
\includegraphics[width=60mm]{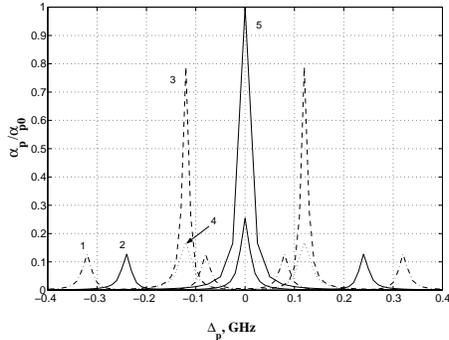}
\vspace{-4mm}
\caption{\label{f3}\footnotesize Absorption index at
homogeneously broadened probe transition
$\left| 1\right\rangle -\left| 4\right\rangle $
 controlled with driving fields.
$\Delta_d= \Delta_c= 0$; 1 -- (dash-dot)
$\Omega_d=120$ MHz, $\Omega_c=200$ MHz; 2 -- (solid)
$\Omega_d=\Omega_c=120$ MHz; 3 -- (dash) $\Omega_d=0$,
$\Omega_c=120$ MHz; 4 -- (dot) $\Omega_d=120$ MHz, $\Omega_c=0$;
5 -- (solid) $\Omega_d=\Omega_c=0$.}
\end{floatingfigure}

We shall illustrate major outcomes of the paper with numerical
simulations addressed to the conditions of the experiment
\cite{Wlg}.
Transitions $\left|1\right\rangle - \left|3
\right\rangle - \left|2\right\rangle - \left|4 \right\rangle -
\left|1\right\rangle$ in Fig.~\ref{f1}a are attributed to those of the
sodium dimers $Na_2$: $X'\Sigma_g^+(v"=0,J"=45)-
A'\Sigma^+_u(6,45)(\lambda_d= 655$ $ nm) - X^1\Sigma_g^+(14,45)
(\lambda_{32}= 756$ $ nm) - B^1\Pi_u(5,45) (\lambda_c= 532$ $nm) -
X'\Sigma_g^+(0,45) (\lambda_p= 480$ $nm)$. The Doppler widths of
the transition at wavelength $ \lambda _ p = 480 $ $ nm $ at the
temperature about 450 C is $2D_{p}\approx 1.7$ GHz. Then the
Boltzmann's population of the level $ \left|2\right\rangle $
makes about 1.5\% from that of the level $\left|1\right\rangle$.
The following relaxation parameters are used: $ \Gamma _ {4} =
\Gamma _ {3} =120$, $ \Gamma _ {2} = \Gamma _ {1}= 20 $, $ \gamma _
{42} = 5 $, $ \gamma _ {41} = 10 $, $ \gamma _ {32} = 4 $, $
\gamma _ {31} = 7 $, $ \Gamma _ {12} = 20 $, $ \Gamma _ {34} = 120 $,
$ \Gamma_ {23} = \Gamma _ {c} = \Gamma _ {d} = \Gamma _ {p} = 70$,
(all in $ 10 ^ {6} $ $ s^ {-1} $). For numerical simulation we have
used the full set of the equations from \cite{Spi,Mys}, accounting
for various relaxation transitions
and not zero population of the level $\left|2\right\rangle$.

We begin our discussion with the case when the Doppler-broadening
is not included. Performing numerical calculation for Eq. (6), we
depict the absorption profiles in Fig.~\ref{f2}. From Fig.~\ref{f2} a and b,
it is easy to find the typical EIT and absorption spectrum with
$\left| \Omega _d\right| =0$ (Fig.~\ref{f2}a) and $\left| \Omega
_c\right| =0$ (Fig.~\ref{f2}b). Under increasing Rabi frequency $\left| \Omega _c\right| $ ($%
\left| \Omega _d\right| $), the four-peak spectrum forms, when the
driving fields become comparable to or larger than the spectral width $%
\Gamma _{p}$ of the atomic transition. The emergence of the
four-peak spectrum is attributed to the fact that levels $\left| 1\right\rangle $ and $%
\left| 4\right\rangle $ have been driven into two dressed states
(Fig.~\ref{f1}). As $\left| \Omega _d\right| $ approaches $\left|
\Omega _c\right| $, the four-peak spectrum degenerates into a
three-peak spectrum since two dressed-state transitions have the
same resonant frequency and contribute to the same central
component. As already known, a three-peak spectrum occurs in
resonance fluorescence (RF) in the strong field limit of a
two-level system (for a review, see \cite{Rau}). Even though the
dressed-state diagrams are similar in the two cases, the
four-peak spectrum can not appear in RF because in the case of RF
the dynamic Stark splitting for the two doublets must be the
same, while in our system the two doublet splittings are
controlled separately by different driving field intensity and
frequency. We would like to stress that the multiple EIT windows
can appear simultaneously with the multiple absorption peaks. In
Fig.~\ref{f2}c, the absorption amplitude $A$ at $\Delta _p=0$ is
plotted against the Rabi frequencies for two resonant driving
fields. An important feature is that the maximum absorption
occurs under the condition that the two Rabi frequencies are
equal, which is due to the fact that the two dressed-state
transitions simultaneously contribute to the central component.
This result may provide a way to measure the field intensity as
well as atomic parameters.

Fig.\ref{f3}   depicts absorption spectrum for the molecules with the
velocity projection on the propagation direction of the coupled
fields $\rm v=0$. The plots illustrate dependencies, described by the
equations (\ref{rt2}) - (\ref{rt4}). The appearance of the
four-peak spectrum is attributed to the fact that levels $\left|
1\right\rangle $ and $ \left| 4\right\rangle $ have been driven
into two dressed states. Direct computing shows that frequency integral
absorption for the plot 3 is the same as for 5, whereas for the
other plots it is about 0.5 of that for the graph 5. This is due
to strong saturation of population difference at the transition
$\left| 1\right\rangle $ -- $ \left| 3\right\rangle $.
\section{Coherence induced resonances in Doppler-broadened medium of sodium dimers}
In the previous sections, Doppler broadening has not been
accounted for. In order to consider this effect, Doppler shifts
must be introduced in the three interacting field resonance
detunings in Eq. (\ref{alpha}).
\begin{figure}[!h]
  \centering
\includegraphics[width=.48\textwidth]{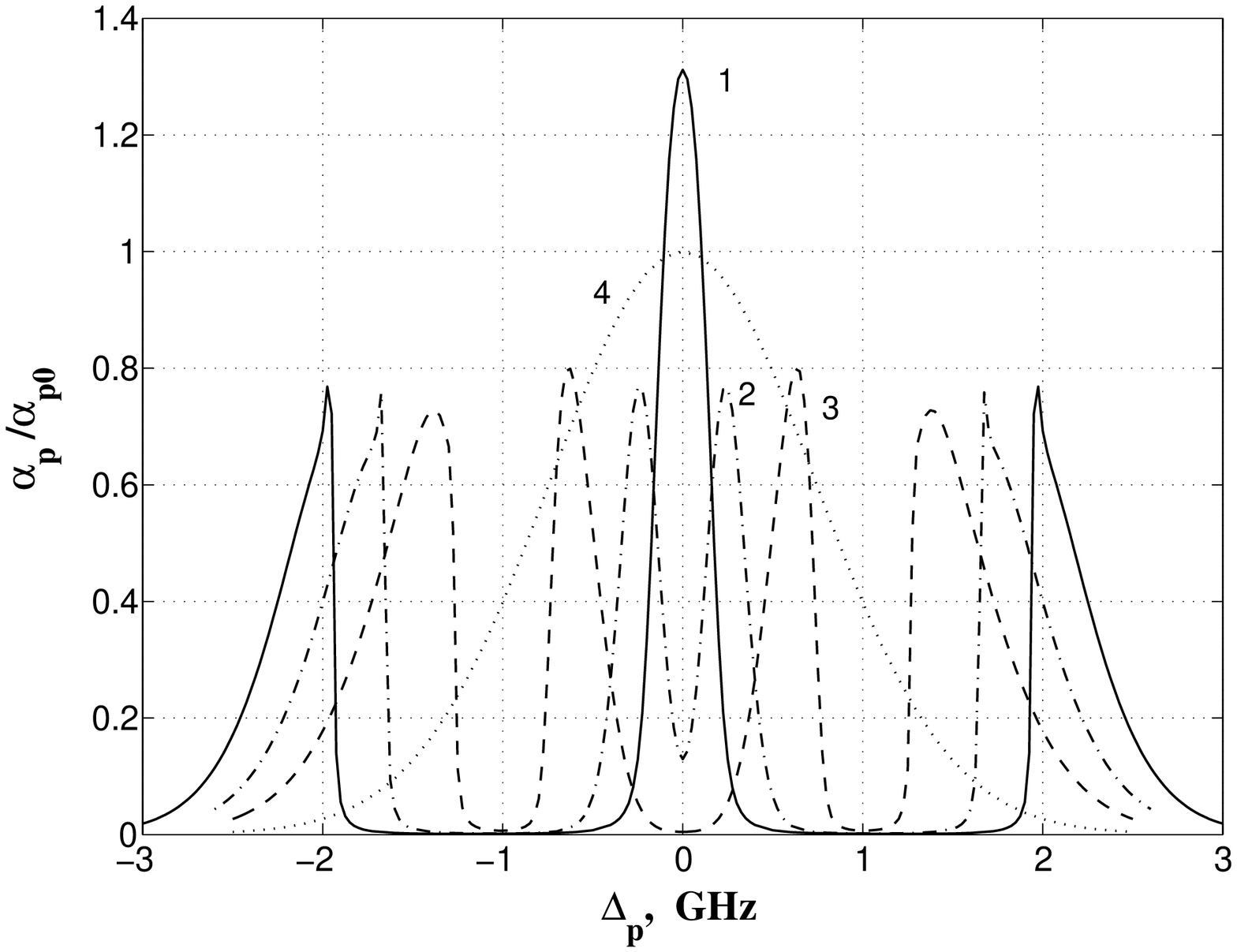}\hfill
\includegraphics[width=.48\textwidth]{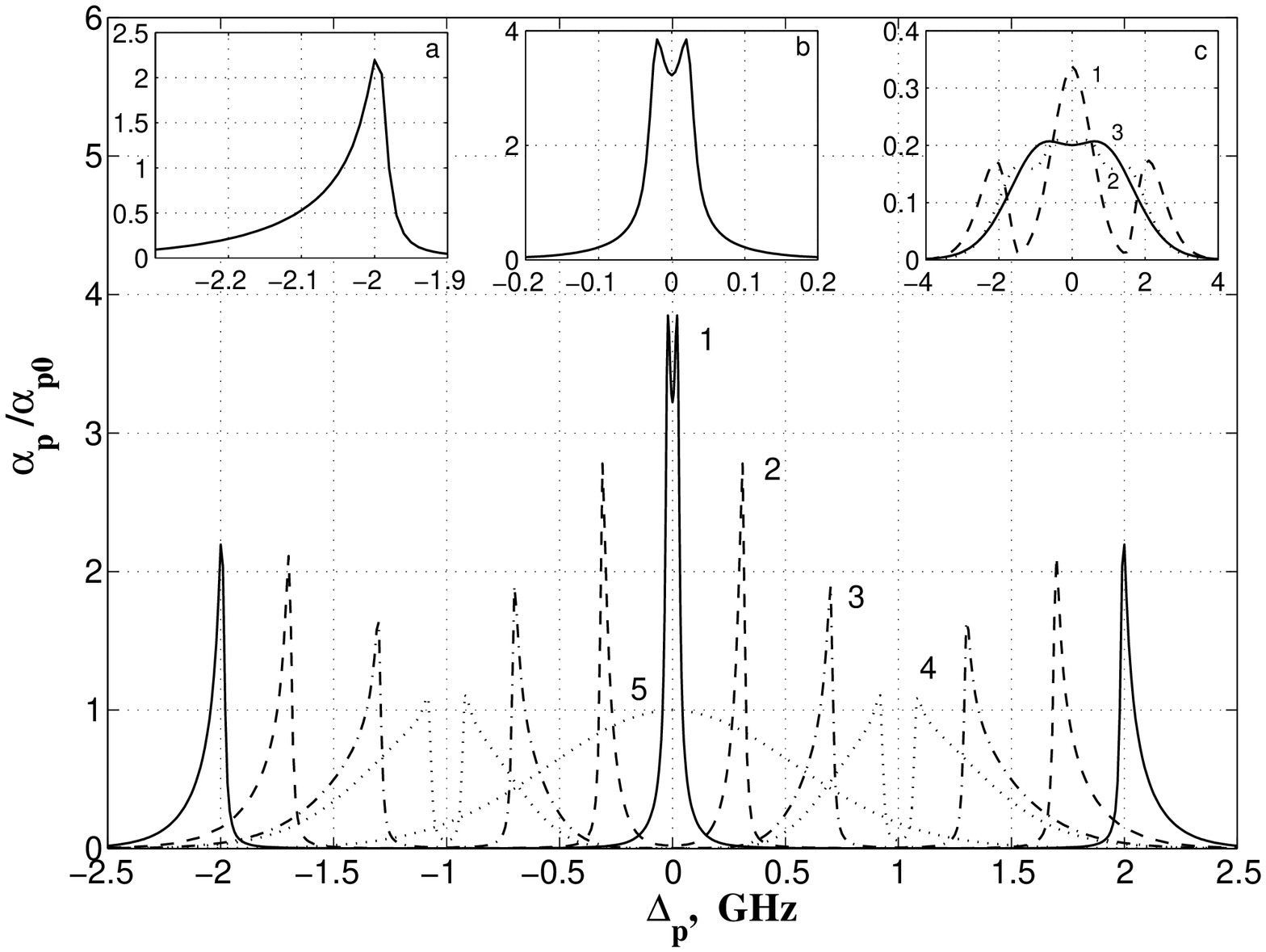}
\parbox{.48\textwidth}{\vspace{-12mm}\caption{\footnotesize Absorption spectrum at inhomogeneously
broadened probe transition $\left| 1\right\rangle -\left| 4\right\rangle $
(the scheme Fig.\protect\ref{f1}(a)) controlled with two driving field. All fields are
co-propagating. $\Delta_d=\Delta_c=0$, $\Omega_c=1$ GHz, $\Omega_d$ is as
follows: 1 -- (solid) $\Omega_d=1$ GHz; 2 -- (dash) $\Omega_d=0.7$ GHz; 3 -- (dot)
$\Omega_d=0.3$ GHz; 4 -- (solid)-- $\Omega_c=\Omega_d=0$. }\label{f4}}\hfill
\parbox{.48\textwidth}{\caption{\footnotesize Compensation of residual Doppler
broadening and sub-Doppler structures at the probe transition $\left| 1\right\rangle
-\left| 4\right\rangle $ (the scheme Fig.\protect\ref{f1}b) coherently driven with two strong fields. Main
plot -- all fields are co-propagating, 1 -- 3 --- all parameters and notations are the
same as in Fig.\protect\ref{f4}. 4 -- (dash-dot): $\Omega_d=0.075$ GHz. Insets (a) and (b) are
corresponding zoomed peaks. Inset (c) --- same as plot 1 but: 1 --- only field $c$ is
counter propagating; 2 --- only field $d$ is counter propagating; 3 --- both driving
fields are counter-propagating in respect to the probe field.}\label{f5}}
\end{figure}
Below we will
proceed with analysis of the absorption spectrum under effects of
Doppler-broadening.
Figure~\ref{f4} displays same dependencies, but for  velocity averaged
absorption index and co-propagating probe and driving fields.
While number of peaks and their positions are identical to the
previous case, the specific feature of the inhomogeneous
broadening is that plot 1 displays enhanced absorption in the
central component despite the fact that integral intensity of
this graph is about 2 times less than that for the plot 4 (due to
the strong saturation of the populations at the transition $\left|
1\right\rangle $ -- $ \left| 3\right\rangle $). In absence of
driving fields only small fraction of the molecules (about
$\Gamma_p/D_{p}\approx 10^{-3}$) can be coupled by probe field.
Overlap of the two dressed transitions gives rise to increased
amount of the molecules coupled with the probe field and
consequently to enhanced absorption as compared to not perturbed
absorption in the center of Doppler broadened transition.

As it was shown in \cite{Feok}, resonance requirements for
molecules at velocity $\rm v$ and at
$max\{|\Omega_d|^2,\Delta_d|^2\}\gg k_du$ ($u$ is thermal
velocity, $k_d$ - wave number) are  described by the equation
\begin{equation}
\Delta_p-{\bf k}_p{\bf v}=\delta_{1,2}-M_{1,2}{\bf k}_d{\bf
v}.\label{df}
\end{equation}
The equation indicates that cancellation of Doppler shifts  is
possible even at  $k_p\neq k_d$, if $k_p < k_d$. This is due to
the interplay of the Doppler and ac Stark shifts giving rise to
variation of $M_{1}$ in the interval 0...1/2, whereas $M_2$ -- in
the interval 1...1/2. Possible enhancements of cross-section of
optical processes through concurrent coupling of molecules from
wide velocity interval by means of the above outlined approach
are shown in Fig.~\ref{f5}. The frequencies of the probe and $E_d$ fields
are interchanged (see Fig.~\ref{f1}b), all other relaxation parameters
remain the same. Plot 1 displays enhanced sub-Doppler resonances
with the FWHM comparable with the natural linewidth (see insets
(a),(b))for co-propagating wave. Dramatic change as compared to
similar graphs in Fig.~\ref{f5} is explicitly seen. The other plots show
evolution of the structures with variation of the intensity of
one of the driving field. As plot 4 indicates, the sub-Doppler
structures exist only under certain ratio of the driving field
intensities. Even more dramatic change occurs while driving
fields (especially shorter wavelength one) are counter
propagating (see inset (c)).
\section{Conclusion}
In conclusion, we have explored spectral properties of the
molecular transitions in the case that both upper and lower levels
are coupled to other atomic levels and Doppler effects play
an important role. The features of the four-peak or three-peak
spectra induced by the driving fields are investigated. Despite
the dominated Doppler-broadening, two or more peaks may possess
the sub-Doppler resolusion. A crucial role of the ratio of the
frequencies of the coupled transition is shown. With manipulation
the detunings and/or intensities of two coupling fields, one can
improve the spectral resolution. The enhanced absorption in the
sub-Doppler peak is shown to be realized, whereas the peak can be
induced in the center of the inhomogeneously broadened transition.
The predicted effects are attributed to the quantum coherence and
interference processes, while frequency-correlation properties of
multi-photon processes experience substantial modification with
the growth of the driving fields intensities, which leads to
corresponding dramatic changes of the role of Doppler effects.

The similar sub-Doppler technique can also be implemented for
other atomic coherence effects, such as EIT, LWI and for enhancing
various resonant nonlinear optical phenomena in inhomogeneously
broadened media. In one of our previous papers \cite{Dong2}, the
inversionless amplification has been considered with the
important feature that incoherent excitation to the upper level
is not necessary in the same scheme as considered in this letter.
Similar scheme is often used in quantum optics. Most recently,
electromagnetically induced absorption was studied in a similar
four-level system \cite{Tach}. We believe that, by utilizing this
sub-Doppler technique, the above effects in inhomengeneously
broadened media can be further enhanced and manipulated.

We want to stress that in optically thick media the processes similar to optical
parametric amplification and involved in the schemes Fig.~\ref{f1}a,b may play
a crucial role, that imposes dramatic consequences on the features of the output
probe signal \cite{OPA}.

\section{Acknowledgements}
One of the authors, Jin-Yue Gao, would like to acknowledge the support from NSF in
China, the research Fund for the Doctoral Program of Higher Education in China,
Education Department of China, and DFG in Germany. A. K. Popov acknowledges support
from the Russian (grant 99-02-39003) and Krasnoyarsk Regional Foundations for Basic
Research, from the Center on Fundamental Natural Sciences at St. Petersburg
University (grant 97-5.2-61) and INTAS (INTAS-99-19). He gratefully acknowledge
support of his visit in Germany from Institute of Quantum Optics of Hannover
University and along with Jin-Yue Gao would like to thank B. Wellegehausen for
stimulating discussions. The authors thank S. A. Myslivets for valuable assistance in
numerical simulation.



\begin{thebibliography}{99}
\bibitem{Rau} S. G. Rautian and A. M. Shalagin, {\it Kinetic
Problems  of  Nonlinear Spectroscopy\/} (Elsevier, Amsterdam,
1991).
\bibitem{Vved} A. K. Popov, {\it Introduction  in Nonlinear Spectroscopy\/},
Novosibirsk, Nauka, 1983 (in Russ).
\bibitem{Feok} The principles of sub-Doppler nonlinear laser spectroscopy
and their interrelation with modification of
frequency-correlation properties of multi-photon processes with
the growth of the intensity of the driving field were considered
in: T. Ya. Popova, A. K. Popov, S. G. Rautian and A. A. Feoktistov, Sov.
Phys. JETP,{\bf 30}, 243 (1970) [Translated from Zh.  Eksp. Teor.
Fiz., {\bf 57},  444 (1969)], quant-ph/0005081.
\bibitem{Let} V. S. Letokhov and  V. P. Chebotaev, {\it
Nonlinear  Spectroscopy\/}, (Springer- Verlag, 1977).
\bibitem{Coh} C. Cohen-Tannoudji, et al.,  Opt. Comm.  {\bf 27}, 71  (1978); S.
Reynaud, et al., Phys. Rev. Lett. {\bf 42}, 756 (1979); S.
Reynaud, et al., Opt. Comm. {\bf 42}, 39 (1982).
\bibitem{Tal} A. K. Popov and L. N. Talashkevich,  Optics Comm., {\bf 28},315
(1979);  A. K. Popov and V. M. Shalaev, Optics Comm., {\bf 35},189
(1980); A. K. Popov and V. M. Shalaev, Opt. Spectrosc.,{\bf 49},
336 (1981) [Transl. from Opt.Spektrosk. {\bf 49}, 617 (1980)];
A. K. Popov and V. M. Shalaev, Sov. J. Quant. Electr. {\bf 12},
289 (1982) [Transl. from Kvant. Electr. {\bf 9}, 488 (1982)].
\bibitem{S.E. Harris}  S. E. Harris, J. E. Field, and A. Imamo\v glu, Phys.
Rev. Lett. {\bf 64}, 1107 (1990); K.-J. Boller, A. Imamo\v glu, and S. E.
Harris, Phys. Rev. Lett. {\bf 66}, 2593(1991); J. E. Field, K. H. Hahn, and
S. E. Harris, Phys. Rev. Lett. {\bf 67}, 3062 (1991).
\bibitem{AWI1}The effect of AWI was proposed and numerically illustrated for three
level $V$  scheme of neon transitions in: T. Ya. Popova,
A. K. Popov, Zhurn.Prikl. Spektrosk., {\bf 12}, No 6, 989, (1970) [Translated in
Engl.:\ Journ. Appl. Spectr {\bf 12}, No 6,  734, (1970), quant-ph/0005047 ;
T. Ya. Popova, A. K. Popov, Izv.Vysh. Uchebn. Zaved., Fizika No 11, 38, (1970)
[Translated in Engl.: Soviet Phys. Journ. {\bf 13}, No 11, 1435, (1970)],
quant-ph/0005049,  also
analyzed in \cite{Vved}; it was realized in the experiments making use of
the neon transition: I. M. Beterov, Cand. Sci., Novosibirsk, Dec. 1970;
a review of the early theoretical and experimental studies
of nonlinear interference effects in Doppler broadened media with special
consideration of AWI can be found also in: A. K. Popov and S. G. Rautian, Proc.  SPIE,
{\bf 2798} ("Coherent Phenomena and Amplification without Inversion", A. V. Andreev,
O. Kocharovskaya and P. Mandel, Editors), 49 (1996), quant-ph/0005114;
A. K. Popov, Bull. Russ. Acad. Sci., Physics, {\bf 60}, 927 (1996)(Allerton Press,
N. Y.) [Transl. from: Izvestiya RAN, ser. Fiz., {\bf 60}, 92 (1996)], quant-ph/0005108.
\bibitem{AWI2} O. Kocharovskaya and Ya. I. Khanin, JETP Lett. {\bf 48}, 630 (1998);
S.E. Harris, Phys. Rev. Lett. {\bf 62},
1033 (1989); M. O. Scully, S. Y. Zhu and A. Gavrielides, Phys.
Rev. Lett. {\bf 62}, 2813 (1989); J. Y. Gao, et. al, Opt. Commun.
{\bf 93}, 323 (1992).
\bibitem{Kuch} A. K. Popov and B. Wellegehausen, Laser Physics {\bf 6},
364 (1996); A. K. Popov, V. M. Kuchin, and S. A. Myslivets, JETP
{\bf 86}, 244 (1998).
\bibitem{Spi} A. K. Popov,  Proc. SPIE {\bf 3485}, 252 (1998), quant-ph/0005118;
also in : {\it Physics of Vibrations} {\bf6}(1), 50 (1998) ed. F. V. Bunkin,
Allerton Press, Inc., New York, USA.
\bibitem{M. O. Scully}  M. O. Scully, Phys. Rev. Lett. {\bf 67}, 1855 (1991);
M. Fleischhauer, C. H. Keitel, and M. O. Scully, Chang Su, B. T.
Ulrich, Shi-Yao Zhu, Phys. Rev. A {\bf 46}, 1468 (1992);
Han-Zhuang Zhang, Xiu-Zhen Guo, Jin-Yue Gao, Yun Jiang, Guang-Xu
Jin, Z. Phys. D {\bf 42}, 83 (1997).
\bibitem{Rev} M. Scully, Phys. Rep., {\bf 219}, 191 (1992);
O. Kocharovskaya, ibid, 175 (1992); B. G. Levi, Physics Today, 17
(May 1992); P. Mandel, Contemp.  Phys., {\bf 34}, 235 (1993);
M. O. Scully and M. Fleischhauer, Science, {\bf 63}, 337 (1994);
S. E. Harris, Physics Today, {\bf 52}, 36 (1997).
\bibitem{Vemuri}  G. Vemuri, G. . Agarwal, and B. D. Nageswara Rao, Phys.
Rev. A {\bf 53}, 2842 (1996).
\bibitem{YiFu Zhu}  Yifu Zhu and T. N. Wasserlauf, Phys. Rev. A {\bf 54},
3653 (1996).
\bibitem{Wang DZ}  De-Zhong, Wang and Jin-Yue Gao Phys. Rev. A {\bf 52},
3201 (1996); G. S. Agarwal and W. Harshawardhan, Phys. Rev. Lett. {\bf 77},
1039 (1996); G. Vemuri, G. S. Agarwal, and B. D. Nageswara Rao, Phys. Rev. A
{\bf 54}, 3695 (1996).
\bibitem{Dong1}  Po Dong and Jin-Yue Gao, Phys. Lett. A {\bf 265}, 52 (2000).
\bibitem{Popov}  A. S. Baev and A. K. Popov, JETP Lett. {\bf 67}, 1018 (1998);
{\bf 69}, 110 (1999); A. K. Popov and V. M. Shalaev, Phys. Rev. A
{\bf 59}, R946 (1999); A. S. Baev, A. K. Popov, S. A. Myslivets, and
V. M. Shalaev, Proc.  SPIE,{\bf 3736}, 279 (1999); A. K. Popov and A. S. Baev,
Phys. Rev. A., August 1 (2000) (to be published), quant-ph/0005089.
\bibitem{NIEF} Nonlinear interference effects in quantum transitions as
the source of amplification without inversion and various feasibilities
manipulating lineshape of optical
transitions with control fields both for discreete and continuum
states, including formation of sign-changing profiles and
transparency and  accounting for Doppler broadening and
collisions, were considered in details in early publications:
G. E. Notkin, S. G. Rautian, and A. A. Feoktistov, JETP {\bf 25},
1112 (1967);
T. Ya. Popova, A. K. Popov, S.  G. Rautian, and R. I. Sokolovskii, JETP {\bf 30},
466 (1970) [Translated from  Zh. Eksp. Teor. Fiz. {\bf 57} 850,
(1969)], quant-ph/0005094; also in \cite{Vved}(for review see refs.
\cite{Rau,AWI1,Spi} above.)
\bibitem{Mys} A. K. Popov and S. A. Myslivets,   Quantum Electron. {\bf 27}(11),
1004 (1997); {\bf 28}(2), 185 (1998) [Transl. from Kvant.
Elektron. {\bf 24}(11), 1033 (1997); {\bf 25}(2),192 (1998)].
\bibitem{Wlg} A. Apolonskii, S. Baluschev, U. Hinze, E. Tiemann and
B.~Wellegehausen, Appl. Phys. B {\bf 64}, 435 (1997); S. Babin,
E. V. Podivilov, D. A. Shapiro, U. Hinze, E. Tiemann and
B. Wellegehausen, Phys. Rev. A {\bf 59}, 1355 (1999).
\bibitem{Dong2}  Po Dong, Weining Man and Jin-Yue Gao, Phys. Lett. A {\bf 265}, 43 (2000).
\bibitem{Tach}  A. V. Taichenachev, A. M. Tumaikin, and V. I. Yudin, Phys. Rev.
A {\bf 61}, 011802 (2000).
\bibitem{OPA} T. F. George and A. K. Popov, submitted; A. K. Popov, S. A. Myslivets, and
T. F. George, submitted.
\end{thebibliography}
\end{document}